\documentclass[aps,prl,twocolumn,showpacs,groupedaddress]{revtex4}
\usepackage{amssymb}
\usepackage{amsmath}
\usepackage{graphicx}

\input{tcilatex}

\begin{document}

\title{Separability Criteria for Arbitrary Quantum Systems}
\author{Chang-shui Yu\thanks{%
quaninformation@sina.com}, He-shan Song}
\affiliation{Department of Physics, Dalian University of Technology,\\
Dalian 116024, China}
\date{\today }

\begin{abstract}
The purpose of this paper is to obtain a sufficient and necessary condition
as a criteria to test whether an arbitrary multipartite state is entangled
or not. Based on the tensor expression of a multipartite pure state, the
paper show that a state is separable iff $\left\vert \mathbf{C}(\rho
)\right\vert $ =0 for pure states and iff $C(\rho )$ vanishes for mixed
states.
\end{abstract}

\maketitle

\section{Introduction}

Entanglement is an essential ingredient in quantum information and the
central feature of quantum mechanics which distingishes a quanum system from
its classical counterpart. As an important physical resource, it is also
widely applied to a lot of quanum information processing(QIP): quantum
computation [1], quantum cryptography [2], quantum teleportation [3],
quantum dense coding [4] and so on.

Entanglement arises only if there has been interactions between the
subsystems of a multipartite system from physics or only if the quantum
state is nonseparable or nonfactorized from mathematics. Even though a lot
of efforts have been made on how to tell whether a given quantum state is
entangled (separable) or not, \ only bipartite entanglement measures
[6,7,8,9] as separability criteria have been for the most part well
understood. Even though the separability criteria for pure states [10,11,12]
are versatile and complex, there does not exist a unified one; A general
formulation of multipartite mixed states is relatively lacking and remains
an open problem.

Recently, Reference [5] has presented a sufficient and necessary condition
for separability of tripartite qubit systems by arranging $a_{ijk}$s of a
pure state $\left\vert \psi \right\rangle _{ABC}=\sum a_{ijk}\left\vert
i\right\rangle _{A}\left\vert j\right\rangle _{B}\left\vert k\right\rangle
_{C}$ as a three-order tensor in $2\times 2\times 2$ dimension. The
introduction of the new skill has provided an effective way to generalize
the criteria to tripartite states in arbitrary dimension and to multipartite
quantum systems.

In this Letter, \ we continue Ref.[5] to give out a general formulation of
separability criterion for arbitrary quantum systems. The paper is organized
as follows: Firstly, we present a sufficient and necessary condition of
separability for tripartite pure states. Secondly, we generalize the
condition to the case of mixed states. Lastly, we generalize our result to
multipartite quantum systems.

\section{Separability criterion for tripartite pure states}

We begin with the definition of our tensors. Unlike the previous definition
of tensors, for convenience, all the quantities with indices, such as $%
T_{ij\cdot \cdot \cdot k}$ and so on, are called tensors here. The number of
the indices is called the order of the tensor. Therefore, the set of all
one-order tensor is the set of vectors, and the set of all two-order tensors
is the one of matrices. Three-order tensors $T_{ijk}$ are matrices (vectors)
if any one (two) of their three indices is (are) fixed.

The elements of a three-order tensor can be arranged at the node of the grid
in three-dimensional Hilbert space, such as the tensor $T_{ijk}$ with $%
i,j,k=0,1,2$ shown in figure 1. From geometry, every fixed index corresponds
to a group of parallel planes which are perpendicular to the vector which
the fixed index corresponds to. The planes corresponding to different fixed
indices are mutually perpendicular.

\begin{figure}[tbp]
\includegraphics[width=7.5cm]{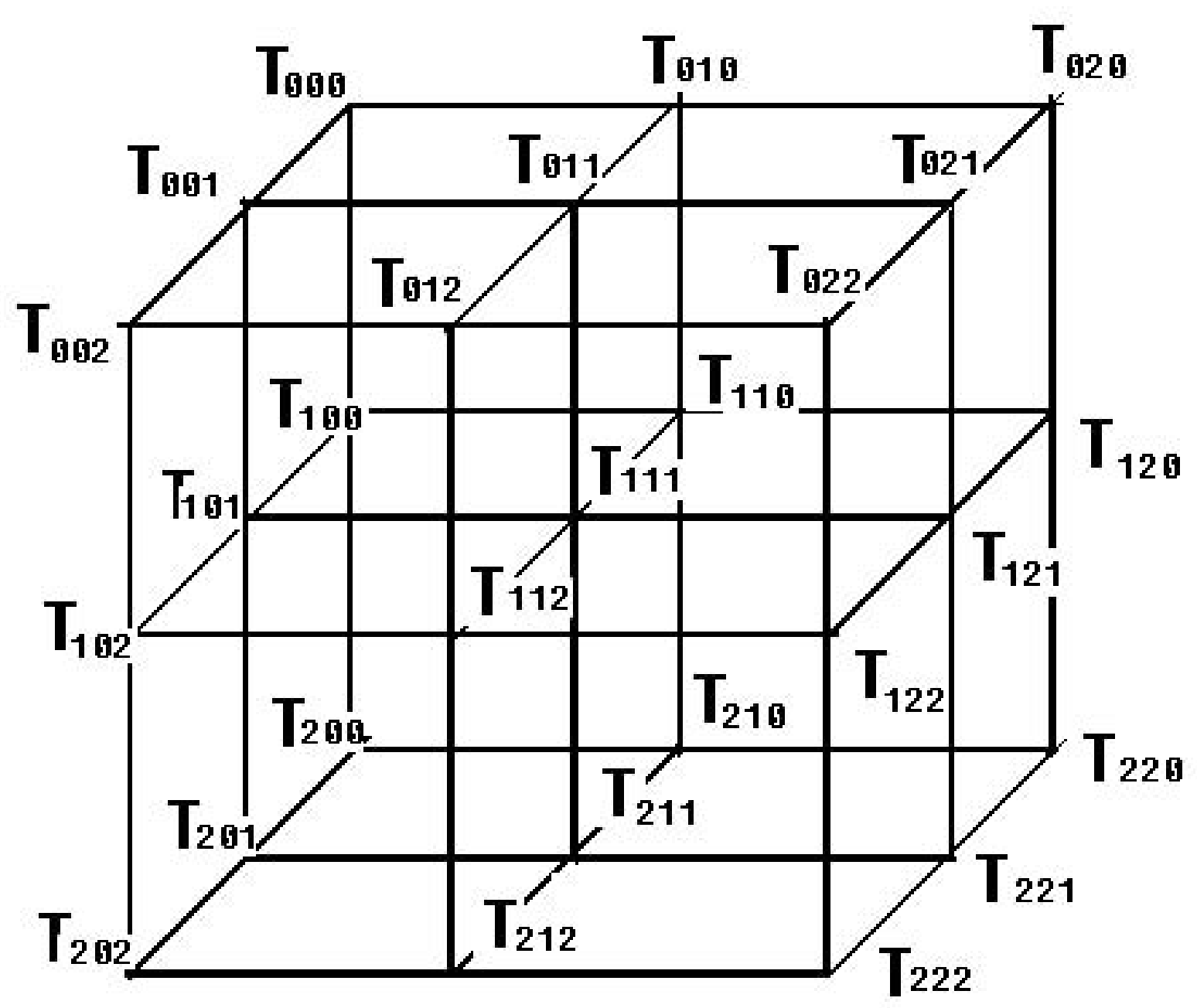}
\caption{Three-order tensor of the coefficients of a tripartite pure state.}
\label{1}
\end{figure}

\bigskip\ \textbf{Definition.-}Let $T_{ijk}$ is a three-order tensor in $%
n_{1}\times n_{2}\times n_{3}$ dimension with $i=0,1,\cdot \cdot \cdot
,n_{1}-1$, $j=0,1,\cdot \cdot \cdot ,n_{2}-1$ and $k=0,1,\cdot \cdot \cdot
,n_{3}-1$. $T_{i^{\prime }j^{\prime }k^{\prime }}^{\prime }$ is any
sub-tensor in $m\times m\times m$ dimension by selecting any $m$ planes from
every group of parallel ones corresponding to three different indices from $%
T_{ijk}$. Then $M_{\alpha \beta \gamma }$=$f(T_{i^{\prime }j^{\prime
}k^{\prime }}^{\prime })$ is called the $m$th compound tensor denoted by $%
C_{m}(T)$ defined in $\left( 
\begin{array}{c}
n_{1} \\ 
m%
\end{array}%
\right) \times \left( 
\begin{array}{c}
n_{2} \\ 
m%
\end{array}%
\right) \times \left( 
\begin{array}{c}
n_{3} \\ 
m%
\end{array}%
\right) $ dimension, where $f(x)$ is a given function of the tensor $x$. For
a two-order tensor $x$, $f(x)=\det x$ is the determinant of the matrix $x.$%
For higher-order tensors, what $f(x)$ denotes is still an open problem.
However, to the intent of this paper, we can give out a $f(x)$ for $%
C_{2}(T_{ijk})$ which is effective enough to characterize the separability
of a tripartite quantum state. For $C_{2}(T)$, every sub-tensor $%
T_{i^{\prime }j^{\prime }k^{\prime }}^{\prime }$ is defined in $2\times
2\times 2$ dimension which corresponds to a tensor cube [5]. So we can
employ 
\begin{equation*}
f(T_{i^{\prime }j^{\prime }k^{\prime }}^{\prime })=\left\vert \mathbf{C}%
\right\vert =\frac{1}{\sqrt{3}}\sqrt{\sum_{\alpha }(C^{\alpha })^{2}},
\end{equation*}%
which is introduced in Ref.[5], where $C^{\alpha }=\left\vert \left(
T_{ijk}\right) ^{\prime }s^{\alpha }T_{ijk}\right\vert $ with $s^{1}=-\sigma
_{y}\otimes \sigma _{y}\otimes I$, $s^{2}=-\sigma _{y}\otimes I\otimes
\sigma _{y}$, $s^{3}=-I\otimes \sigma _{y}\otimes \sigma _{y}$, $%
s^{4}=-Iv\otimes \sigma _{y}\otimes \sigma _{y}$, $s^{5}=-\sigma _{y}\otimes
Iv\otimes \sigma _{y}$, $s^{6}=-\sigma _{y}\otimes \sigma _{y}\otimes Iv)$,
here $\sigma _{y}=\left( 
\begin{array}{cc}
0 & -i \\ 
i & 0%
\end{array}%
\right) $, $I=\left( 
\begin{array}{cc}
1 & 0 \\ 
0 & 1%
\end{array}%
\right) $, and $Iv=\left( 
\begin{array}{cc}
0 & 1 \\ 
1 & 0%
\end{array}%
\right) $.

Analogous to Ref.[5], considering any a tripartite pure state $\left\vert
\psi \right\rangle _{ABC}=\sum a_{ijk}\left\vert i\right\rangle
_{A}\left\vert j\right\rangle _{B}\left\vert k\right\rangle _{C}$ with $%
i=0,1,\cdot \cdot \cdot ,n_{1}-1$, $j=0,1,\cdot \cdot \cdot ,n_{2}-1$ and $%
k=0,1,\cdot \cdot \cdot ,n_{3}-1$, if arranging the coefficients of the
state as a tensor denoted by $A_{ijk}$ (every coefficient corresponds to a
node of the grid), every line in the grid corresponds to a vector and every
plane corresponds to a matrix. One can easily find that, the tripartite
state is\ fully separable iff all the parallel vectors are linear relevant.
A necessary and sufficient condition for this which is easily proved is that
the second compound tensor $C_{2}(A_{ijk})$ must be zero. Namely, every
element of the tensor must be zero. Consider the state $\left\vert \psi
\right\rangle _{ABC}$ written in vector notation $\left\vert \psi
\right\rangle _{ABC}$ $=$($a_{000},a_{001},\cdot \cdot \cdot
,a_{00n_{3}-1},a_{010},\cdot \cdot \cdot ,a_{n_{1}-1n_{2}-1n_{3}-1})^{\prime
}$, one can obtain an equivalent expression of above relation, i.e.%
\begin{equation*}
\left\vert \mathbf{C}_{\alpha \beta \gamma }(\psi )\right\vert =\frac{1}{%
\sqrt{3}}\sqrt{\sum_{p}(C_{\alpha \beta \gamma }^{p}(\psi ))^{2}}=0
\end{equation*}%
holds for any $\alpha $, $\beta $ and $\gamma $, where $C_{\alpha \beta
\gamma }^{p}=\left\vert \left\langle \psi _{ABC}\right\vert s_{\alpha \beta
\gamma }^{p}\left\vert \psi _{ABC}\right\rangle \right\vert $ with $%
s_{\alpha \beta \gamma }^{1}=-L_{\alpha }\otimes L_{\beta }\otimes I_{\gamma
}$, $s_{\alpha \beta \gamma }^{2}=-L_{\alpha }\otimes I_{\beta }\otimes
L_{\gamma }$, $s_{\alpha \beta \gamma }^{3}=-I_{\alpha }\otimes L_{\beta
}\otimes L_{\gamma }$, $s_{\alpha \beta \gamma }^{4}=-\left\vert L_{\alpha
}\right\vert \otimes L_{\beta }\otimes L_{\gamma }$, $s_{\alpha \beta \gamma
}^{5}=-L_{\alpha }\otimes \left\vert L_{\beta }\right\vert \otimes L_{\gamma
}$, $s_{\alpha \beta \gamma }^{6}=-L_{\alpha }\otimes L_{\beta }\otimes
\left\vert L_{\gamma }\right\vert $, here $L_{\alpha }$, $L_{\beta }$ and $%
L_{\gamma }$ are the generators of $SO(n_{1})$, $SO(n_{2})$ and $SO(n_{3})$,
respectively; $I_{\alpha }$, $I_{\beta }$ and $I_{\gamma }$ are the unit
matrices in $n_{1}$, $n_{2}$ and $n_{3}$ dimension, respectively; with $%
\alpha =1,2,\cdot \cdot \cdot ,\frac{n_{1}(n_{1}-1)}{2}$, $\beta =1,2,\cdot
\cdot \cdot ,\frac{n_{2}(n_{2}-1)}{2}$ and $\gamma $ $=1,2,\cdot \cdot \cdot
,\frac{n_{3}(n_{3}-1)}{2}$. $\left\vert M\right\vert $ denotes the modulus
of the elements of the matrix $M$.

Then we can construct a new vector $\mathbf{C}$=$\underset{\alpha \beta
\gamma }{\oplus }\mathbf{C}_{\alpha \beta \gamma }$ and employ the length of
the vector 
\begin{equation*}
\left\vert \mathbf{C(}\psi )\right\vert =\sqrt{\underset{\alpha \beta \gamma 
}{\sum }\left\vert \mathbf{C}_{\alpha \beta \gamma }\mathbf{(}\psi
)\right\vert ^{2}}=\frac{1}{\sqrt{3}}\sqrt{\sum_{p}\underset{\alpha \beta
\gamma }{\sum }(C_{\alpha \beta \gamma }^{p}\mathbf{(}\psi ))^{2}}
\end{equation*}%
as the criterion of separability.

\section{Separability criterion for tripartite mixed states}

The tripartite mixed states $\rho =\sum\limits_{k=1}\omega _{k}\left\vert
\psi ^{k}\right\rangle \left\langle \psi ^{k}\right\vert $ can be written in
matrix notation as $\rho =\Psi W\Psi ^{\dagger }$, where $W$ is a diagonal
matrix with $W_{kk}=\omega _{k}$, the columns of \ the matrix $\Psi $
correspond to the vectors $\psi ^{k}$. Consider the eigenvalue
decomposition, $\rho =\Phi M\Phi ^{\dagger }$, where $M$ is a diagonal
matrix whose diagonal elements are the eigenvalues of $\rho $, and $\Phi $
is a unitary matrix whose columns are the eigenvectors of $\rho $. From
Ref.[], one can get $\Psi W^{1/2}=\Phi M^{1/2}T$, where $T$ is a
Right-unitary matrix. The tripartite mixed states are fully separable iff
there exist a decomposition such that $\psi ^{k}$ for every $k$ is fully
separable. The entanglement measure of formation can be defined as the
infimum of the average $\left\vert \boldsymbol{C}(\psi ^{k})\right\vert $.
Namely, $C(\rho )=\inf \sum\limits_{k}\omega _{k}\left\vert \boldsymbol{C}%
(\psi ^{k})\right\vert $, if $C(\rho )$ is assigned as the entanglement
measure for tripartite mixed states. Therefore, for any a decomposition 
\begin{equation*}
\rho =\sum\limits_{k=1}\omega _{k}\left\vert \psi ^{k}\right\rangle
\left\langle \psi ^{k}\right\vert ,
\end{equation*}%
one can get%
\begin{eqnarray}
C(\rho ) &=&\inf \sum\limits_{k}\omega _{k}\left\vert \boldsymbol{C}(\psi
^{k})\right\vert  \notag \\
&=&\inf \sum\limits_{k}\omega _{k}\frac{1}{\sqrt{3}}\sqrt{\sum_{p}\underset{%
\alpha \beta \gamma }{\sum }\left\vert \left\langle (\psi ^{k})^{\ast
}\right\vert s_{\alpha \beta \gamma }^{p}\left\vert \psi ^{k}\right\rangle
\right\vert ^{2}}.  \notag
\end{eqnarray}%
According to the Mincowski inequality%
\begin{equation}
\left( \sum\limits_{i=1}\left( \sum\limits_{k}x_{i}^{k}\right) ^{p}\right)
^{1/p}\leq \sum_{k}\left( \sum\limits_{i=1}\left( x_{i}^{k}\right)
^{p}\right) ^{1/p},\text{ }p>1,
\end{equation}%
one can easily obtained 
\begin{eqnarray}
C(\rho ) &\geq &\inf \frac{1}{\sqrt{3}}\sqrt{\sum_{p}\underset{\alpha \beta
\gamma }{\sum }\left( \sum\limits_{k}\omega _{k}\left\vert \left\langle
(\psi ^{k})^{\ast }\right\vert s_{\alpha \beta \gamma }^{p}\left\vert \psi
^{k}\right\rangle \right\vert \right) ^{2}}  \notag \\
&=&\inf \frac{1}{\sqrt{3}}\sqrt{\sum_{p}\underset{\alpha \beta \gamma }{\sum 
}\left( \sum\limits_{k}\left\vert \Psi ^{T}W^{1/2}s_{\alpha \beta \gamma
}^{p}W^{1/2}\Psi \right\vert _{kk}\right) ^{2}}  \notag \\
&=&\underset{T}{\inf }\frac{1}{\sqrt{3}}\sqrt{\sum_{p}\underset{\alpha \beta
\gamma }{\sum }\left( \sum\limits_{k}\left\vert T^{T}M^{1/2}\Phi
^{T}s_{\alpha \beta \gamma }^{p}\Phi M^{1/2}T\right\vert _{kk}\right) ^{2}} 
\notag \\
&\geq &\underset{T}{\inf }\frac{1}{\sqrt{3}}\sum\limits_{k}\left\vert
T^{T}\left( \sum_{p}\underset{\alpha \beta \gamma }{\sum }z_{\alpha \beta
\gamma }^{p}A_{\alpha \beta \gamma }^{p}\right) T\right\vert _{kk},
\end{eqnarray}%
where $A_{\alpha \beta \gamma }^{p}=M^{1/2}\Phi ^{T}s_{\alpha \beta \gamma
}^{p}\Phi M^{1/2}$ for any $z_{\alpha \beta \gamma }^{p}=y_{\alpha \beta
\gamma }^{p}e^{i\phi }$ with $y_{\alpha \beta \gamma }^{p}>0$, $%
\sum\limits_{\alpha }\left( y_{\alpha \beta \gamma }^{p}\right) ^{2}=1$, and
Cauchy-Schwarz inequality 
\begin{equation}
\left( \sum\limits_{i}x_{i}^{2}\right) ^{1/2}\left(
\sum\limits_{i}y_{i}^{2}\right) ^{1/2}\geqslant \sum\limits_{i}x_{i}y_{i},
\end{equation}%
are applied at the last step. The infimum of equation (3) is given by $%
\underset{z\in \mathbf{C}}{max}\lambda _{1}(z)-\underset{i>1}{\sum }\lambda
_{i}(z)$ with $\lambda _{i}(z)$s are the singular values, in decreasing
order, of the matrix $\underset{p}{\frac{1}{\sqrt{3}}\sum }\underset{\alpha
\beta \gamma }{\sum }z_{\alpha \beta \gamma }^{p}A_{\alpha \beta \gamma
}^{p} $. Therefore, we can express $C(\rho )$ as 
\begin{equation}
C(\rho )=\max \{0,\underset{z\in \mathbf{C}}{max}\lambda _{1}(z)-\underset{%
i>1}{\sum }\lambda _{i}(z)\}.
\end{equation}

It is not difficult to find that $C(\rho )=0$ is a sufficient and necessary
condition of separability for mixed states according to the whole procedure
of derivation.

\section{Separability criterion for multipartite systems}

A general $N$-partite pure states 
\begin{eqnarray}
\left\vert \psi \right\rangle _{AB\cdot \cdot \cdot N}
&=&\sum\limits_{ij\cdot \cdot \cdot k}a_{ij\cdot \cdot \cdot k}\left\vert
ij\cdot \cdot \cdot k\right\rangle ,  \notag \\
i &\in &[0,n_{1}-1],j\in \lbrack 0,n_{2}-1],\cdot \cdot \cdot ,k\in \lbrack
0,n_{N}-1],
\end{eqnarray}%
is separable iff $\left\vert \psi \right\rangle _{AB\cdot \cdot \cdot
N}=\sum\limits_{ij\cdot \cdot \cdot k}a_{ij\cdot \cdot \cdot k}\left\vert
i\right\rangle _{A}\otimes \left\vert j\right\rangle _{B}\otimes \cdot \cdot
\cdot \otimes \left\vert k\right\rangle _{N}$. Analogously, if the
coefficients $a_{ij\cdot \cdot \cdot k}$s are arranged as an $N$-order
tensor, one can easily find that, $\left\vert \psi \right\rangle _{AB\cdot
\cdot \cdot N}$ is separable iff all the vectors which are mutually parallel
are linear relevant. In order to obtain a mathematical rigorous criterion,
we have to redefine matrices $s_{\underset{N}{\underbrace{\alpha \beta \cdot
\cdot \cdot \lambda }}}^{ij}$s as 
\begin{eqnarray}
s_{\underset{N}{\underbrace{\alpha \beta \cdot \cdot \cdot \lambda }}}^{ij}
&=&L_{\alpha }\otimes L_{\beta }\otimes \underset{i}{\underbrace{\left\vert
L_{\gamma }\right\vert \otimes \cdot \cdot \cdot \otimes \left\vert
L_{\delta }\right\vert }}\otimes \underset{N-i-2}{\underbrace{I_{\rho
}\otimes ,\cdot \cdot \cdot ,\otimes I_{\lambda }}}, \\
i &=&0,1,\cdot \cdot \cdot ,N-2,\text{ \ }j=1,2,\cdot \cdot \cdot ,\binom{N}{%
i}\times \binom{N-i}{2},
\end{eqnarray}%
where $L_{x}$ denotes the generators of $SO(n_{p})$, $x=1,2,\cdot \cdot
\cdot ,\frac{n_{p}(n_{p}-1)}{2},$ with $p$ standing for the $p$th subsystem. 
$i$ in above equation (6) states that there are $i$ absolute values of
generators. Note that the order of $L_{\alpha }$, $L_{\beta }$, $\left\vert
L_{\gamma }\right\vert $, $\cdot \cdot \cdot $, $\left\vert L_{\delta
}\right\vert $, $I_{\rho }$, $\cdot \cdot \cdot $, $I_{\lambda }$ in
equation (6) must cover all the permutations with $j$ as a index showing the 
$j$th permutation.

Hence, the criterion for pure states can be expressed as following, if an $N$%
-partite pure state is separable iff 
\begin{equation*}
\left\vert \mathbf{C}(\psi )\right\vert =\sqrt{\sum (C_{\alpha \beta \cdot
\cdot \cdot \gamma }^{ij}(\psi ))^{2}}/\sqrt{\binom{N}{2}}=0,
\end{equation*}%
where the sum is over all the indices and $1/\sqrt{\binom{N}{2}}$ is a
normalized factor.

According to the same procedure of the derivation to Section II, one can
easily obtain the criterion of separability for an $N$-partite mixed state
by testing whether $C(\rho )$ vanishes with 
\begin{equation*}
C(\rho )=\max \{0,\underset{z\in \mathbf{C}}{max}\lambda _{1}(z)-\underset{%
i>1}{\sum }\lambda _{i}(z)\},
\end{equation*}%
where $\lambda _{i}(z)$s are the singular values, in decreasing order, of
the matrix $\underset{ij}{\sum }\underset{\alpha \beta \cdot \cdot \cdot
\gamma }{\sum }z_{\alpha \beta \cdot \cdot \cdot \gamma }^{ij}A_{\alpha
\beta \cdot \cdot \cdot \gamma }^{ij}/\sqrt{\binom{N}{2}}$.

Let us finally note that, our result can also be reduced to Wootters,
concurrence when $N=2$ in 2$\times 2$ dimension or the result in Ref.[5]
owing to the only generator $\sigma _{y}$ of $SO(2)$. Therefore, the final
result presented in this Letter is a general one suitable for arbitrary
quantum systems.

\section{\protect\bigskip Conclusion}

As a summary, in this paper, we generalize the result presented in Ref.[5]
to arbitrary quantum systems. The result in this paper provides a sufficient
and necessary condition as a general criterion to test whether a given
multipartite system is separable or not. The criterion for pure states is
convenient and analytic, but it has to turn to a numerical optimization for
mixed states, however, a simple treatment similar to Ref.[5] is often
enough. The fruitful conclusion similar to [8] is expectable and compensable.

\section{Acknowledgement}

This work was supported by Ministry of Science and Technology, China, under
grant No.2100CCA00700.

\end{document}